\documentstyle[12pt]{article}
\renewcommand{\baselinestretch}{1.5}
\newcommand{\resection}[1]{\setcounter{equation}{0}\section{#1}}
\newcommand{\appsection}{\addtocounter{section}{1} \setcounter{equation}{0}
                         \section*{Appendix \Alph{section}}}
\renewcommand{\theequation}{\thesection.\arabic{equation}}
\newcommand{\EQ}{\begin{equation}}
\newcommand{\EN}{\end{equation}}
\newcommand{\spz}{\hspace{0.7cm}}
\newcommand{\hs}{\hspace{0.1cm}}
\newcommand{\th}{\theta}
\newcommand{\s}{\sigma}
\newcommand{\goto}{\rightarrow}
\newcommand{\be}{\beta}
\newcommand{\k}{\kappa}
\def\zwi{\hspace{0.2 in}}
\begin{document}
\oddsidemargin 5mm
\setcounter{page}{0}
\renewcommand{\thefootnote}{\fnsymbol{footnote}}
\newpage
\setcounter{page}{0}
\begin{titlepage}
\begin{flushright}
ISAS/94/91/EP \\
NORDITA 91/47
\end{flushright}
\vspace{0.5cm}
\begin{center}
{\large {\bf On the S-matrix of the Sub-leading Magnetic Deformation of
the Tricritical Ising Model in Two Dimensions}} \\
\vspace{1.5cm}
{\bf F. Colomo$^1$\footnote{Angelo Della Riccia Foundation's fellow.},
A. Koubek$^2$,
G. Mussardo$^{1}$\footnote{On leave of absence from:
International School for Advanced Studies, Strada Costiera 11, 34014
Trieste, Italy.}} \\
\vspace{0.8cm}
$^1${\em NORDITA, Blegdamsvej 17, DK-2100 Copenhagen \O, Denmark }\\
$^2${\em International School for Advanced Studies, Strada Costiera 11,
34014 Trieste, Italy} \\
\end{center}
\vspace{6mm}
\begin{abstract}
We compute the $S$-matrix of the Tricritical Ising Model perturbed by
the subleading magnetic operator using Smirnov's RSOS reduction of the
Izergin-Korepin model. The massive model contains kink excitations which
interpolate between two degenerate asymmetric vacua.
As a consequence of the different structure of the two vacua, the crossing
symmetry is implemented in a non-trivial way. We use finite-size
techniques to compare our results with the numerical data obtained by the
Truncated Conformal Space Approach and find good agreement.
\end{abstract}
\vspace{5mm}
\centerline{June 1991}
\end{titlepage}

\newpage

\renewcommand{\thefootnote}{\arabic{footnote}}
\setcounter{footnote}{0}
\resection{Introduction}
It has been pointed out by Zamolodchikov that certain deformations of
minimal models of Conformal Field Theories (CFT) produce integrable
massive field theories, which are characterized on mass-shell by a factorizable
$S$-matrix \cite{Zam}. A particularly interesting situation occurs for
the ``Tricritical Ising Model'' (TIM) at the fixed point perturbed by the
operator $\Phi_{2,1}$.  This field has anomalous dimensions
$(\Delta,\overline\Delta)=(\frac{7}{16},\frac{7}{16})$ and is identified
with the sub-leading magnetic operator\footnote{In the Landau-Ginzburg theory,
the TIM represents the universality class of the $\varphi^6$ model. The
operator $\Phi_{2,1}$ is represented by the third-order field $\varphi^3$.},
odd with respect to the $Z_2$ spin-reversal transformation. Hence, this
deformation explicitly breaks the $Z_2$ symmetry of the tricritical point
and the corresponding massive theory can exhibit ``$\Phi^3$-property'',
i.e. the absence of a conserved current of spin 3 and the possibility to form a
bound state through the process $A A \hs\rightarrow \hs A \rightarrow \hs A
A$. The counting argument supports this picture, giving for the
spin of the conserved currents the values s=(1,5,7,11,13) \cite{Zam,LMC}.
The interesting features of this massive field theory have been first outlined
in ref.\hs\cite{LMC} where the model was studied by the ``Truncated Conformal
Space Approach'' (TCSA), proposed in ref.\hs \cite{YZ}. This approach consists
in the diagonalization of the perturbed Hamiltonian
\EQ
H=H_{CFT} + \lambda \int \hs \Phi_{2,1} (x) \hs dx
\EN
on a strip\footnote{We consider in the following only the
case of periodic boundary condition.} of width $R$, truncated at a certain
level of the Hilbert space defined by the conformal field theory at the fixed
point. The lowest energy levels are given in fig.\hs1. The theory presents two
degenerate ground states (which correspond to the minima of the asymmetric
double-well Landau-Ginzburg potential in fig.\hs2) and a single excitation
$B$ of mass $m$ below the threshold at $2m$. The double degeneracy of the
vacuum permits two fundamental kink configurations $\mid K_+\rangle$ and
$\mid K_-\rangle$ and, possibly, bound states thereof. If the two vacua were
related by a symmetry transformation, one would expect a double degeneracy of
the breather-like bound state $\mid B\rangle$ in the infrared regime
$R\rightarrow \infty$. However, the absence of a $Z_2$ symmetry makes it
possible that in this case only one of the two asymptotic states
$\mid K_+ K_-\rangle$ or $\mid K_- K_+\rangle$ is coupled to the bound
state $\mid B\rangle$.

A scattering theory for such model with asymptotic states $\mid K_+\rangle$,
$\mid K_-\rangle$ and $\mid B \rangle$ was first conjectured in
\cite{sublead}. On the other hand, a general framework for the $\Phi_{1,2}$
and $\Phi_{2,1}$ deformations of CFT has been recently proposed in
\cite{Smirnov}. It is based on the RSOS reduction of the Izergin-Korepin
model \cite{IK}. In section 3 we explicitly work out the RSOS-like $S$-matrix
in the case of the TIM perturbed by the subleading magnetic operator. The
scattering theory we obtain is different from the one of
ref.\hs\cite{sublead}, although both of them have the following
features:
\begin{enumerate}
\item the existence of two fundamental kink configurations with mass
$m$;
\item the appearance of only one bound state of the above, with the same mass
$m$.
\end{enumerate}
Hence, both scattering theories give rise to a picture that qualitatively
agrees with the results of ref.\hs\cite{LMC}. Therefore a more detailed
analysis is required in order to decide which of the two theories is
appropriate for the description of the scaling region of the TIM in presence
of a third-order magnetic perturbation. As suggested in ref.\hs\cite{sublead},
possible insight comes from the study of finite-size effects for the
theory defined on the cylinder \cite{Luscher}. These finite-size effects,
which can be directly related to the scattering data, control the exponential
decay of the one-particle energy level to its asymptotic value.

The paper is organized as follows: in sect.\hs 2 we outline
Smirnov's RSOS reduction of the Izergin-Korepin model. In sect.\hs 3 we
present our results on the $S$-matrix of the $\Phi_{2,1}$ perturbation of
TIM. Sect.\hs 4 contains the analysis of the phase shifts and the
asymptotic behaviour of the RSOS $S$-matrix. Sect.\hs5 consists of a short
summary of Zamolodchikov's proposal for the $S$-matrix \cite{sublead}.
This proposal, together with ours, will be checked against numerical data
in sect.\hs 6, where we perform an accurate analysis of the energy
levels obtained from the TCSA using the finite-size theory. Our conclusions
are in sect.\hs7.

\resection{Smirnov's RSOS reduction of the Izergin-Korepin model}

In a recent paper, Smirnov has related the $\Phi_{1,2}$ and $\Phi_{2,1}$
deformations of a CFT to a RSOS projection of the $R$-matrix of the
Izergin-Korepin (IK) model \cite{Smirnov,IK}. His results can be
summarized as follows. The $R$-matrix of the IK model cannot be directly
interpreted as the $S$-matrix of the massive excitations of the perturbed CFT
because, introducing a Hilbert space structure, it does not satisfy the
unitarity requirement. But it is possible to use the quantum $SL(2)_q$
symmetry of the IK model to study the RSOS restriction of the
Hilbert space. This happens when $q$ is a $p$th root of unity.
The RSOS reduction preserves the locality of an invariant set of operators and
yields $S$-matrices which have a sensible physical interpretation. The RSOS
states appearing in the reduced model
\EQ
\mid \beta_1,j_1,k_1, \mid a_1\mid\hs \beta_2,j_2,k_2, \ldots
\mid a_{n-1}\mid \hs \beta_n,j_n,k_n\rangle
\EN
are characterized by their rapidity $\beta_i$, by their type $k$ (which
distinguish the kinks from the breathers), by their $SL(2)_q$ spin $j$ and
by a string of integer numbers $a_i$, satisfying
\EQ
a_i \leq \frac{p-2}{2}\hs\hs , \hspace{6mm}
\mid a_k-1\mid \leq\hs
a_{k+1} \leq \hs\mbox{min}(a_k+1,p-3-a_k) \hs .
\label{limitation}
\EN
In the case of $\Phi_{2,1}$ deformation, using the following parameterization
of the central charge of the original CFT
\EQ
c=1-6\left(\frac{\pi}{\gamma}+\frac{\gamma}{\pi}-2\right) \hs ,
\EN
the $q$-parameter is given by
\EQ
q=\exp(2 i \gamma) \hs .
\EN
For the unitary minimal CFT, $\gamma=\pi\hs \frac{r}{r+1}$ ($r=3,4,\dots $).
It is also convenient to define the quantity
\EQ
\xi=\frac{2}{3} \left(\frac{\pi^2}{2\gamma-\pi}\right) \hs .
\EN
The S-matrix of the RSOS states is given by
\cite{Smirnov}

\newpage

\begin{eqnarray}
\lefteqn{S_{a_{k-1} a_{k+1}}^{a_k a_k'}\left(\beta_k-\beta_{k+1}\right)
=} \nonumber \\
& & \frac{i}{4}\hs S_0(\beta_k-\beta_{k+1})\left[
\left\{\begin{array}{ccc}
1 & a_{k-1} & a_k \\
1 & a_{k+1} & a_k'
\end{array} \right\}_q \right. \hs \hspace{5mm} \nonumber \\
& & \times
\left(\left(\exp\left(\frac{2\pi}{\xi}(\beta_{k+1}-\beta_k)\right)
-1\right)\hs q^{(c_{a_{k+1}}+c_{a_{k-1}}-c_{a_k}-c_{a_k'}+3)/2}\right.
\label{Skink} \\
& &\left. -\left(\exp\left(-\frac{2\pi}{\xi}(\beta_{k+1}-\beta_k)\right)
-1\right)\hs q^{-(c_{a_{k+1}}+c_{a_{k-1}}-c_{a_k}-c_{a_k'}+3)/2}\right)
\nonumber \\
& & \left.
\begin{array}{cc}
 & \\
 &
\end{array}
+ q^{-5/2} (q^3+1) (q^2-1) \delta_{a_k,a_k'}\right].
\nonumber
\end{eqnarray}
Herein, $c_a$ are the Casimir of the representation $a$, $c_a=a(a+1)$
and the expression of the $6j$-symbols is that in ref.\hs\cite{KirRes}.
$S_0(\beta)$ has the following integral representation
\begin{eqnarray}
S_0(\beta) &=& \left(\sinh\frac{\pi}{\xi}(\beta-i\pi)
\sinh\frac{\pi}{\xi}\left(\beta-\frac{2\pi i}{3}\right)\right)^{-1} \nonumber\\
& & \mbox{} \times \exp\left(
-2i \int_0^{\infty} \frac{dx}{x} \frac{\sin \beta x \sinh\frac{\pi
x}{3} \cosh\left(\frac{\pi}{6}-\frac{\xi}{2}\right)x}
{\cosh\frac{\pi x}{2} \sinh\frac{\xi x}{2}}\right) \hs .
\end{eqnarray}

\resection{RSOS $S$-matrix of $\Phi_{2,1}$ perturbed TIM}

In the TIM perturbed by the subleading magnetization operator, $r=4$
and $\xi=\frac{10 \pi}{9}$. From eq.\hs(\ref{limitation}), the only possible
values of $a_i$ are $0$ and $1$ and the one-particle states are the
vectors: $\mid K_{01} \rangle$, $\mid K_{10} \rangle$ and
$\mid K_{11} \rangle$. All of them have the same mass $m$. Notice that the
state $\mid K_{00}\rangle$ is not allowed. A basis for the two-particle
asymptotic states is
\EQ
\mid K_{01} K_{10} \rangle, \hs\hs
\mid K_{01} K_{11} \rangle, \hs\hs
\mid K_{11} K_{11} \rangle, \hs \hs
\mid K_{11} K_{10} \rangle, \hs \hs
\mid K_{10} K_{01} \rangle \hs .
\label{basis}
\EN
The scattering processes are
\begin{eqnarray}
\mid K_{01}(\beta_1) K_{10}(\beta_2) \rangle & = &
S_{00}^{11} (\beta_1-\beta_2) \mid K_{01}(\beta_2) K_{10} (\beta_1)
\rangle
\nonumber \\
\mid K_{01}(\beta_1) K_{11}(\beta_2) \rangle & = &
S_{01}^{11} (\beta_1-\beta_2) \mid K_{01}(\beta_2) K_{11} (\beta_1)
\rangle
\nonumber \\
\mid K_{11}(\beta_1) K_{10} (\beta_2)\rangle & = &
S_{10}^{11} (\beta_1-\beta_2) \mid K_{11} (\beta_2) K_{10} (\beta_1)
\rangle
\label{scattering}\\
\mid K_{11} (\beta_1) K_{11} (\beta_2)\rangle & = &
S_{11}^{11} (\beta_1-\beta_2) \mid K_{11} (\beta_2) K_{11} (\beta_1)
\rangle +
S_{11}^{10}(\beta_1-\beta_2) \mid K_{10} (\beta_2) K_{01}
(\beta_1) \rangle \nonumber \\
\mid K_{10} (\beta_1) K_{01} (\beta_2) \rangle & = &
S_{11}^{00} (\beta_1-\beta_2) \mid K_{10} (\beta_2) K_{01} (\beta_1)
\rangle +
S_{11}^{10}(\beta_1-\beta_2) \mid K_{11} (\beta_2) K_{11}(\beta_1)
\rangle \nonumber
\end{eqnarray}
Explicitly, the above amplitudes are given by

\vspace{10mm}

\begin{picture}(190,40)
\thicklines
\put(60,0){\line(1,1){30}}
\put(60,30){\line(1,-1){30}}
\put(60,15){\makebox(0,0){$0$}}
\put(90,15){\makebox(0,0){$0$}}
\put(75,0){\makebox(0,0){$1$}}
\put(75,30){\makebox(0,0){$1$}}
\put(100,15){\makebox(0,0)[l]{$\displaystyle{\hs =\hs S_{00}^{11}(\beta)
\hs =\hs \frac{i}{2}\hs
S_0(\beta)
\hs\sinh\left(\frac{9}{5} \beta -i \frac{\pi}{5}\right)}$}}
\put(390,15){\makebox(0,0)[l]{(3.3.a)}}
\end{picture}

\vspace{10mm}

\begin{picture}(190,40)
\thicklines
\put(60,0){\line(1,1){30}}
\put(60,30){\line(1,-1){30}}
\put(60,15){\makebox(0,0){$0$}}
\put(90,15){\makebox(0,0){$1$}}
\put(75,0){\makebox(0,0){$1$}}
\put(75,30){\makebox(0,0){$1$}}
\put(100,15){\makebox(0,0)[l]{$\displaystyle{\hs =\hs S_{01}^{11}(\beta)
\hs =\hs -\frac{i}{2}\hs
S_0(\beta)
\hs\sinh\left(\frac{9}{5} \beta +i \frac{\pi}{5}\right)}$}}
\put(390,15){\makebox(0,0)[l]{(3.3.b)}}
\end{picture}

\vspace{10mm}

\begin{picture}(190,40)
\thicklines
\put(60,0){\line(1,1){30}}
\put(60,30){\line(1,-1){30}}
\put(60,15){\makebox(0,0){$1$}}
\put(90,15){\makebox(0,0){$1$}}
\put(75,0){\makebox(0,0){$1$}}
\put(75,30){\makebox(0,0){$1$}}
\put(100,15){\makebox(0,0)[l]{$\displaystyle{\hs =\hs S_{11}^{11}(\beta)
\hs =\hs \frac{i}{2}\hs S_0(\beta) \hs
\frac{\sin\left(\frac{\pi}{5}\right)}{\sin\left(\frac{2\pi}{5}\right)}
\hs\sinh\left(\frac{9}{5} \beta -i \frac{2\pi}{5}\right)}$}}
\put(390,15){\makebox(0,0)[l]{(3.3.c)}}
\end{picture}

\vspace{10mm}

\begin{picture}(190,40)
\thicklines
\put(60,0){\line(1,1){30}}
\put(60,30){\line(1,-1){30}}
\put(60,15){\makebox(0,0){$1$}}
\put(90,15){\makebox(0,0){$1$}}
\put(75,0){\makebox(0,0){$0$}}
\put(75,30){\makebox(0,0){$1$}}
\put(100,15){\makebox(0,0)[l]{$\displaystyle{\hs =\hs S_{11}^{01}(\beta)
\hs =\hs -\frac{i}{2}\hs S_0(\beta) \hs
\left(\frac{\sin\left(\frac{\pi}{5}\right)}{\sin\left(\frac{2\pi}{5}\right)}
\right)^{\frac{1}{2}}
\hs\sinh\left(\frac{9}{5} \beta\right)}$}}
\put(390,15){\makebox(0,0)[l]{(3.3.d)}}
\end{picture}

\vspace{10mm}

\begin{picture}(190,40)
\thicklines
\put(60,0){\line(1,1){30}}
\put(60,30){\line(1,-1){30}}
\put(60,15){\makebox(0,0){$1$}}
\put(90,15){\makebox(0,0){$1$}}
\put(75,0){\makebox(0,0){$0$}}
\put(75,30){\makebox(0,0){$0$}}
\put(100,15){\makebox(0,0)[l]{$\displaystyle{\hs =\hs S_{11}^{00}(\beta)
\hs =\hs -\frac{i}{2}\hs S_0(\beta) \hs
\frac{\sin\left(\frac{\pi}{5}\right)}{\sin\left(\frac{2\pi}{5}\right)}
\hs\sinh\left(\frac{9}{5} \beta +i \frac{2\pi}{5}\right)}$}}
\put(390,15){\makebox(0,0)[l]{(3.3.e)}}
\end{picture}

\vspace{10mm}

\setcounter{equation}{3}
\noindent

The function $S_0(\beta)$ can actually be computed.
It is given by
\begin{eqnarray}
S_0(\beta) &=& -\left(\sinh\frac{9}{10}(\beta-i\pi)
\sinh\frac{9}{10}\left(\beta-\frac{2\pi i}{3}\right)\right)^{-1} \nonumber\\
& & \mbox{} \times w\left(\beta,-\frac{1}{5}\right)
w\left(\beta,+\frac{1}{10}\right)
w\left(\beta,\frac{3}{10}\right) \\
& & \mbox{} \times
t\left(\beta,\frac{2}{9}\right)
t\left(\beta,-\frac{8}{9}\right) t\left(\beta,\frac{7}{9}\right)
t\left(\beta,-\frac{1}{9}\right) \nonumber\hs ,
\end{eqnarray}
where
\[
w(\beta,x)=\frac{\sinh\left(\frac{9}{10} \beta +i\pi x\right)}
{\sinh\left(\frac{9}{10} \beta -i\pi x\right)}\hs ;
\]
\[
t(\beta,x)=\frac{\sinh\hs \frac{1}{2}(\beta +i\pi x)}
{\sinh\hs \frac{1}{2}(\beta -i\pi x)} \hs .
\]

\vspace{5mm}

\noindent
It is easy to check the unitarity equations:
\begin{eqnarray}
& & S_{11}^{00}(\beta)\hs S_{11}^{00}(-\beta)+
S_{11}^{01}(\beta)\hs S_{11}^{10}(-\beta)=1 \hs ;\nonumber \\
& & S_{11}^{10}(\beta)\hs S_{11}^{01}(-\beta)+
S_{11}^{11}(\beta)\hs S_{11}^{11}(-\beta)=1 \hs ;\nonumber \\
& & S_{11}^{10}(\beta)\hs S_{11}^{00}(-\beta)+
S_{11}^{11}(\beta)\hs S_{11}^{10}(-\beta)=0 \hs ;\label{unitarity}\\
& & S_{10}^{11}(\beta) \hs S_{10}^{11}(-\beta)=1 \hs ;\nonumber \\
& & S_{00}^{11}(\beta) \hs S_{00}^{11}(-\beta)=1 \nonumber \hs .
\end{eqnarray}
An interesting property of this $S$ matrices is that the crossing
symmetry occurs in a non-trivial way, i.e.
\begin{eqnarray}
S_{11}^{11}(i\pi-\beta) & = & S_{11}^{11}(\beta) \hs ;\nonumber \\
S_{11}^{00}(i\pi-\beta) & = & a^2 \hs S_{00}^{11}(\beta)
\hs ; \label{crossing}\\
S_{11}^{01}(i\pi-\beta) & = & a \hs S_{01}^{11}(\beta) \hs ;\nonumber
\end{eqnarray}
where
\EQ
a=-\left(\frac{s\left(\frac{1}{5}\right)}
{s\left(\frac{2}{5}\right)}\right)^{\frac{1}{2}} \hs ,
\label{asymmetry}
\EN
and $s(x)\equiv \sin(\pi x)$.

The above crossing-symmetry relations may be seen as due to a
non-trivial charge conjugation operator (see also \cite{Mussardo}).
In most cases, the charge conjugation is implemented trivially, i.e.
with $a=\pm 1$ in eq.\hs(\ref{crossing}). Here, the asymmetric Landau-Ginzburg
potential distinguishes between the two vacua and gives rise to the value
(\ref{asymmetry}).

The amplitudes (3.3) are periodic along the imaginary axis of $\beta$
with period $10\hs\pi i$. The whole structure of poles and zeros is quite
rich. On the physical sheet, $0\leq {\rm Im} \hs\beta\leq i \pi$, the poles
of the $S$-matrix are located at $\beta=\frac{2\pi i}{3} $ and
$\beta=\frac{i\pi}{3}$ (fig.\hs3). The first pole corresponds to a bound
state in the direct channel while the second one is the singularity
due to the particle exchanged in the crossed process. The residues at
$\beta=\frac{2\pi i}{3}$ are given by
\begin{eqnarray}
r_1 & =& \mbox{Res}_{\beta=\frac{2\pi i}{3}} \hs\hs S_{00}^{11}(\beta) = 0
\hs; \nonumber \\
r_2 & =& \mbox{Res}_{\beta=\frac{2\pi i}{3}} \hs
\hs S_{01}^{11}(\beta) = i \hs
\left(\frac{s\left(\frac{2}{5}\right)}{s\left(\frac{1}{5}\right)}\right)^2
\hs \omega \hs ;\nonumber \\
r_3 & = & \mbox{Res}_{\beta=\frac{2\pi i}{3}} \hs\hs S_{11}^{11}(\beta) =
i\hs \omega \hs ;\label{resRSOS1}\\
r_4 & =& \mbox{Res}_{\beta=\frac{2\pi i}{3}} \hs
\hs S_{11}^{01}(\beta) = i\hs
\left(\frac{s\left(\frac{2}{5}\right)}{s\left(\frac{1}{5}\right)}
\right)^{\frac{1}{2}}
\hs \omega \hs ;
\nonumber \\
r_5 & =& \mbox{Res}_{\beta=\frac{2\pi i}{3}} \hs
\hs S_{11}^{00}(\beta) = i \hs
\frac{s\left(\frac{2}{5}\right)}{s\left(\frac{1}{5}\right)}
\hs \omega \hs ;\nonumber
\end{eqnarray}
where
\EQ
\omega =\frac{5}{9} \hs \frac{
s\left(\frac{1}{5}\right)
s\left(\frac{1}{10}\right)
s\left(\frac{4}{9}\right)
s\left(\frac{1}{9}\right)
s^2\left(\frac{5}{18}\right)}
{s\left(\frac{3}{10}\right)
s\left(\frac{1}{18}\right)
s\left(\frac{7}{18}\right)
s^2\left(\frac{2}{9}\right)} \hs\hs .
\EN

\vspace{3mm}
\noindent
Their numerical values are collected in Table 1.

In the amplitude $S_{00}^{11}$ there is no bound state in the direct
channel but only the singularity coming from to the state $\mid K_{11}\rangle$
exchanged in the $t$-channel. This is easily seen from Fig.\hs 4
where we stretch the original amplitudes along the vertical direction
($s$-channel) and along the horizontal one ($t$-channel). Since the
state $\mid K_{00} \rangle$ is not physical, the residue in the
direct channel is zero. In the amplitude $S_{01}^{11}$ we have
the bound state $\mid K_{01}\rangle$ in the direct channel and the
singularity due to $\mid K_{11}\rangle$ in the crossed channel. In
$S_{11}^{11}$, the state $\mid K_{11}\rangle$ appears as a bound state
in both channels. In $S_{11}^{01}$ the situation is reversed with respect
to that of $S_{01}^{11}$, as it should be from the crossing symmetry property
(\ref{crossing}): the state $\mid K_{11}\rangle$ appears in the
$t$-channel and $\mid K_{01}\rangle$ in the direct channel. Finally, in
$S_{11}^{00}$ there is the bound state $\mid K_{11}\rangle$ in the
direct channel but the residue on the $t$-channel pole is zero, again
because $\mid K_{00}\rangle$ is unphysical. This situation is, of
course, that obtained by applying crossing to $S_{00}^{11}$.

\resection{Energy levels, phase shifts and generalized statistics}

The one-particle line {\em a} of fig.\hs (1.a) corresponds to the state
$\mid K_{11}\rangle$. This energy level is not doubly degenerate
because the state $\mid K_{00}\rangle$ is forbidden by the RSOS selection
rules, eq.\hs(\ref{limitation}). With periodic boundary conditions,
the kink states $\mid K_{01}\rangle $ and $\mid K_{10} \rangle $ are
projected out and $\mid K_{11} \rangle $ is the only one-particle state that
can appear in the spectrum.

Consider the threshold line. On a strip with periodic boundary
conditions we expect this energy level to be doubly degenerate.
This because the states $\mid K_{01}K_{10} \rangle $ and
$\mid K_{10}K_{01} \rangle $ are identified under such boundary conditions
whereas the state $\mid K_{11}K_{11} \rangle $ remains distinct. The conformal
operators creating these states in the u.v. limit are
$\Phi_{\frac{7}{16},\frac{7}{16}}$ and $\Phi_{\frac{6}{10},\frac{6}{10}}$,
respectively (see lines marked {\em b} and {\em c} in fig.\hs1.a). We have
checked that for large values of $R$ these two lines approach each other
faster than $1/r$ (as would be the case if the lowest line {\em b} was
the only threshold line and {\em c} a line of momentum). However the truncation
effects already present in this region prevents us from showing that they
really approach each other exponentially\footnote{Consider, for
instance, figure 10 of ref.\hs\cite{LMC} (the case of low-temperature phase of
thermal perturbation of the TIM). One observes that the onset of the
exponential approach of degenerate excited levels usually occurs quite far
from the value of $R$ at which the two ground state energies coincide. As we
discuss in sect\hs 6, the situation of the subleading magnetic deformation of
the TIM is even worse from a numerical point of view, because the anomalous
dimension of the subleading magnetic operator is approximately $1/2$. In
this case, it is therefore possible that the onset of the exponential approach
of the higher levels takes place in a region of $R$ strongly dominated by
truncation effects.}.
In the following we assume that ideally these two levels are exponentially
degenerate in the infrared limit.

The identification of the threshold lines described above holds only for the
static configuration of two kinks with zero relative momentum. The
situation is indeed different for the lines of momentum which approach
the threshold. In fact, the $S$-matrix acting on $\mid K_{01}K_{10}
\rangle $ can only produce $\mid K_{01}K_{10} \rangle $ as final state
whereas $\mid K_{10}K_{01} \rangle $ and $\mid K_{11}K_{11} \rangle $
can mix through the processes of eq.\hs (\ref{scattering}).
In order to determine the pattern of the energy levels obtained from TCSA
and to relate the scattering processes to the data of the original unperturbed
CFT (along the line suggested in \cite{SM}), we would need a higher-level Bethe
ansatz technique. This is because our actual situation deals with kink-like
excitations in contrast to that of ref.\hs\cite{SM} which considers only
diagonal, breather-like $S$ matrices. The Bethe-ansatz technique gets quite
complicated in the case of a $S$-matrix with kink excitations. For the moment,
it has been applied in few cases \cite{RTBA1,RTBA2}. In the light of these
difficulties, we prefer here not to pursue such a program and instead
concentrate on some properties of the phase shifts and generalized statistics
which arises for kink excitations.

For real values of $\beta$, the amplitudes $S_{00}^{11}(\beta)$ and
$S_{01}^{11}(\beta)$ are numbers of modulus 1. It is therefore convenient to
define the following phase shifts
\begin{eqnarray}
S_{00}^{11}(\beta) & \equiv & e^{2 i \delta_{0}(\beta)} \hs;
\label{phaseshift}\\
S_{01}^{11}(\beta) & \equiv & e^{2 i \delta_{1}(\beta)} \hs .\nonumber
\end{eqnarray}
The non-diagonal sector of the scattering processes is characterized
by the $2\times 2$ symmetric $S$-matrix
\EQ
\left(
\begin{array}{ll}
S_{11}^{11}(\beta) & S_{11}^{01}(\beta) \\
S_{11}^{01}(\beta) & S_{11}^{00}(\beta)
\end{array}
\right)
\label{nondiag} \hs .
\EN

\vspace{3mm}
\noindent
We can define the corresponding phase shifts by diagonalizing the matrix
(\ref{nondiag}). The eigenvalues turn out to be the same functions in
(\ref{phaseshift}),
\EQ
\left(
\begin{array}{cc}
e^{2i\delta_0(\beta)} & 0\\
0 & e^{2i\delta_1(\beta)}
\end{array}
\right)
\label{diag} \hs .
\EN
The phase shifts, for positive values of $\beta$, are shown in fig.\hs5.
Asymptotically, they have the following limits
\begin{eqnarray}
\lim_{\beta\rightarrow\pm\infty} \hs e^{2i \delta_0(\beta)} & = &
e^{\pm\frac{6 \pi i}{5}} \hs ;\\
\lim_{\beta\rightarrow\pm\infty} \hs e^{2 i \delta_1(\beta)} & = &
e^{\pm\frac{3\pi i}{5}}\nonumber \hs .
\end{eqnarray}
There is a striking difference between the two phase shifts: while
$\delta_1(\beta)$ is a monotonic decreasing function, starting from
its value at zero energy $\delta_1(0)=\frac{\pi}{2}$,
$\delta_{0}(\beta)$ shows a maximum for $\beta \sim \frac{\pi}{3}$ and then
decreases to its asymptotic value $\frac{3 \pi}{5}$. Its values are always
larger that $\delta_0(0)=\frac{\pi}{2}$. Such different behaviour of the phase
shifts is related to the presence of a zero very close to the real axis
in the amplitude $e^{2i\delta_0 (\beta)}$, i.e. at $\beta=i\frac{\pi}{9}$.
This zero competes with the pole at $\beta=i\frac{\pi}{3}$ in creating a
maximum in the phase shift. Similar behaviour also occurs in non-relativistic
cases \cite{Schiff} and in the case of breather-like $S$ matrices which
contains zeros \cite{ColMus}. The presence of such a zero is deeply related to
the absence of the pole in the s-channel of the amplitude
$e^{2 i\delta_0(\beta)}$. For the amplitude $e^{2 i\delta_1(\beta)}$, the zero
is located at $\beta=\frac{4 \pi i}{9}$ (between the two poles) and therefore
its contribution to the phase shift is damped with respect to that one
coming from the poles. The net result is a monotonic decreasing phase shift.

Coming back to the $2\times 2$ $S$-matrix of eq.\hs(\ref{nondiag}), a
basis of eigenvectors is given by
\begin{eqnarray}
\mid \phi_1(\beta_1)\phi_1(\beta_2)\rangle & = & A(\beta_{12}) \left(
\mid K_{11}(\beta_1) K_{11}(\beta_2)\rangle + \chi_1(\beta_{12}) \hs
\mid K_{10}(\beta_1) K_{01}(\beta_2) \rangle \right) \hs; \\
\mid \phi_2(\beta_1)\phi_2(\beta_2)\rangle & = & A(\beta_{12}) \left(
\mid K_{11}(\beta_1) K_{11}(\beta_2)\rangle + \chi_2(\beta_{12}) \hs
\mid K_{10}(\beta_1) K_{01}(\beta_2) \rangle \right) \hs .\nonumber
\end{eqnarray}
where $A(\beta_{12})$ is a normalization factor. In the asymptotic regime
$\beta\rightarrow \infty$
\begin{eqnarray}
\chi_1 & = &-\hs
\frac{e^{-\frac{2\pi i}{5}}\hs\left(a^2 + e^{\frac{6\pi i}{5}}\right)}{a}
\\
\chi_2 & = & -\hs
\frac{e^{-\frac{2\pi i}{5}}\hs \left(a^2 + e^{\frac{3\pi i}{5}}\right)}{a}
\hs \hs, \nonumber
\end{eqnarray}
and the probability $P_{1001}$ to find a state $\mid K_{10}K_{01}\rangle$
in the vector $\mid \phi_2\phi_2\rangle$ w.r.t. the probability $P_{1111}$ to
find a state $\mid K_{11}K_{11}\rangle$ is given by the golden ratio
\EQ
\frac{P_{1001}}{P_{1111}}\hs =\frac{1}{a^2}=\hs 2\cos\left(\frac{\pi}{5}\right)
\hs .
\EN
For the state $\mid \phi_1\phi_1\rangle$, we have
\EQ
\frac{P_{1001}}{P_{1111}}=\hs a^2=\hs
\frac{1}{2\cos\left(\frac{\pi}{5}\right)}
\hs .
\EN
The ``kinks'' $\phi_1$ and $\phi_2$ have the generalized bilinear commutation
relation \cite{Swieca,KT,Smirnov2}
\EQ
\phi_i(t,x) \phi_j (t,y) =\phi_j(t,y) \phi_i(t,x)\hs e^{2\pi i s_{ij}
\epsilon(x-y)} \hs.
\EN
The generalized ``spin'' $s_{ij}$ is a parameter related to the
asymptotic behaviour of the $S$-matrix. A consistent
assignment is given by
\begin{eqnarray}
s_{11} & = & \frac{3}{5} =\hs \frac{\delta_0(\infty)}{\pi}\nonumber
\hs\hs;\\
s_{12} & = & 0 \hs\hs;\\
s_{22} & = & \frac{3}{10} =\hs \frac{\delta_1(\infty)}{\pi} \hs\hs.\nonumber
\end{eqnarray}
The implications of these generalized statistics will be discussed elsewhere.
Here we only notice that, interesting enough, the previous monodromy
properties are those of the chiral field $\Psi=\Phi_{\frac{6}{10},0}$ of
the original CFT of the TIM. The operator product expansion of $\Psi$ with
itself reads
\EQ
\Psi(z) \Psi(0) = \frac{1}{z^{\frac{6}{5}}} \hs {\bf 1} +
\frac{C_{\Psi,\Psi,\Psi}}{z^{\frac{3}{5}}} \hs \Psi(0) + \ldots
\label{OPE}
\EN
where $C_{\Psi,\Psi,\Psi}$ is the structure constant of the OPE algebra.
Moving $z$ around the origin, $z\rightarrow e^{2 \pi i} z$, the
phase acquired from the first term on the right hand side of (\ref{OPE})
comes from the conformal dimension of the operator $\Psi$ itself. In contrast,
the phase obtained from the second term is due to the insertion of an
additional operator $\Psi$. A similar structure appears in the scattering
processes of the ``kinks'' $\phi_i$: in the amplitude of the kink $\phi_1$
there is no bound state in the s-channel (corresponding to the
``identity term'' in (\ref{OPE})) whereas in the amplitude of $\phi_2$ a kink
can be created as a bound state for $\beta=\frac{2\pi i}{3}$
(corresponding to the ``$\Psi$ term'' in (\ref{OPE})). In the
ultraviolet limit, the fields $\phi_i$ should give rise to the operator
$\Psi(z)$, similarly to the case analyzed in \cite{Smirnov2}. The
actual proof requires the analysis of the form factors and will be given
elsewhere.

\resection{Zamolodchikov's $S$-matrix for $\Phi_{2,1}$ perturbed TIM}

The problem of finding a theoretical explanation for the energy levels of
the $\Phi_{2,1}$ perturbed TIM was first discussed in a remarkable paper by
Zamolodchikov \cite{sublead}. In his notation, the one particle states are
given by
\EQ
\mid K_+ \rangle, \hs \hs
\mid K_- \rangle, \hs \hs
\mid B  \rangle \hs ,
\EN
which we can identify with our $\mid K_{10}\rangle$, $\mid K_{01}\rangle$
and $\mid K_{11}\rangle$, respectively.
The two-particle amplitudes of the scattering processes were defined in
\cite{sublead} to be
\begin{eqnarray}
\mid B (\beta_1) B (\beta_2)\rangle & = &
a (\beta_1-\beta_2) \mid B (\beta_2) B (\beta_1)
\rangle +
b(\beta_1-\beta_2) \mid K_+ (\beta_2) K_-
(\beta_1) \rangle \nonumber \\
\mid K_-(\beta_1) B (\beta_2)\rangle  & = &
c (\beta_1-\beta_2) \mid K_- (\beta_2) B (\beta_1)
\rangle
\nonumber \\
\mid B (\beta_1) K_+ (\beta_2)\rangle & = &
c (\beta_1-\beta_2) \mid B (\beta_2) K_+ (\beta_1)
\rangle \label{scattering2}
\\
\mid K_-(\beta_1) K_+(\beta_2) \rangle &  = &
d (\beta_1-\beta_2) \mid K_-(\beta_2) K_+ (\beta_1)
\rangle
\nonumber \\
\mid K_+ (\beta_1) K_- (\beta_2) \rangle & = &
e (\beta_1-\beta_2) \mid K_+ (\beta_2) K_- (\beta_1)
\rangle +
b(\beta_1-\beta_2) \mid B (\beta_2) B(\beta_1)
\rangle \nonumber
\end{eqnarray}

They are in correspondence with those of eq.\hs(\ref{scattering}) if we
make the following assignments
\begin{eqnarray}
a(\beta) & \rightarrow & S_{11}^{11}(\beta) \hs;\nonumber \\
b(\beta) & \rightarrow & S_{11}^{10}(\beta) \hs;\nonumber \\
c(\beta) & \rightarrow & S_{01}^{11}(\beta) \hs;\\
d(\beta) & \rightarrow & S_{00}^{11}(\beta) \hs;\nonumber \\
e(\beta) & \rightarrow & S_{11}^{00}(\beta) \hs. \nonumber \hs .
\end{eqnarray}
In order to solve the Yang-Baxter equations which ensure the factorization
of the scattering processes, Zamolodchikov noticed that the above amplitudes
coincide with the definitions of the Boltzmann weights of the ``Hard
Square Lattice Gas''. Therefore, he borrowed Baxter's solution
\cite{Baxter} in the case where it reduces to trigonometric form
\begin{eqnarray}
a(\beta) & =& \frac{\sin\left(\frac{2\pi}{5}+\lambda \beta\right)}
{\sin\left(\frac{2\pi}{5}\right)} \hs R(\beta) \hs\hs;\nonumber \\
b(\beta) & =& e^{\delta \beta}\hs \frac{\sin\left(\lambda \beta\right)}
{\left[\sin\left(\frac{2\pi}{5}\right)
\sin\left(\frac{\pi}{5}\right)\right]^{\frac{1}{2}}} \hs R(\beta) \hs\hs;
\nonumber \\
c(\beta) & = & e^{-\delta \beta}\hs
\frac{\sin\left(\frac{\pi}{5}-\lambda \beta\right)}
{\sin\left(\frac{\pi}{5}\right)} \hs R(\beta) \hs\hs;\\
d(\beta) & =& e^{-2\delta\beta} \hs
\frac{\sin\left(\frac{\pi}{5}+\lambda \beta\right)}
{\sin\left(\frac{\pi}{5}\right)} \hs R(\beta) \hs\hs;\nonumber \\
e(\beta) & =& e^{2\delta\beta} \hs
\frac{\sin\left(\frac{2\pi}{5}-\lambda \beta\right)}
{\sin\left(\frac{2\pi}{5}\right)} \hs R(\beta) \nonumber \hs\hs .
\end{eqnarray}
Here $\delta$ and $\lambda$ are arbitrary parameters and $R(\beta)$ is
an arbitrary function. In order to fix completely the amplitudes,
Zamolodchikov imposed the following requirements:
\begin{enumerate}
\item the unitarity conditions, eqs.(\ref{unitarity});
\item the absence of a pole in the direct channel of the amplitude
$d(\beta)$;
\item crossing symmetry, implemented in the following form
\begin{eqnarray}
a(\beta) & = & a(i\pi -\beta) \hs ;\nonumber \\
b(\beta) & = & c(i\pi -\beta) \hs ;\label{cross2}\\
d(\beta) & = & e(i\pi -\beta) \hs .\nonumber
\end{eqnarray}
\end{enumerate}
The final form of the $S$ matrices is given by
\begin{eqnarray}
a(\beta) & =& e^{-2i\pi \delta} \hs
\frac{\sin\left(\frac{2\pi-6i\beta}{5}\right)
\sin\left(\frac{3\pi+6i\beta}{5}\right)}
{\sin\left(\frac{\pi-6i\beta}{5}\right)
\sin\left(\frac{2\pi+6i\beta}{5}\right)}\hs\hs;\nonumber \\
b(\beta) & =& e^{-\delta(i\pi- \beta)} \hs
\frac{\sin\left(\frac{6i\beta}{5}\right)
\sin\left(\frac{3\pi+6i\beta}{5}\right)}
{\sin\left(\frac{\pi-6i\beta}{5}\right)
\sin\left(\frac{2\pi+6i\beta}{5}\right)}\hs\hs;\nonumber \\
c(\beta) & = & e^{-\delta \beta} \hs
\frac{\sin\left(\frac{\pi+6i\beta}{5}\right)
\sin\left(\frac{3\pi+6i\beta}{5}\right)}
{\sin\left(\frac{\pi-6i\beta}{5}\right)
\sin\left(\frac{2\pi+6i\beta}{5}\right)} \hs\hs;\\
d(\beta) & =& e^{-2\delta\beta} \hs
\frac{\sin\left(\frac{3\pi+6i\beta}{5}\right)}
{\sin\left(\frac{2\pi+6i\beta}{5}\right)}\hs\hs;\nonumber \\
e(\beta) & =& e^{-2\delta(i\pi-\beta)} \hs
\frac{\sin\left(\frac{3\pi+6i\beta}{5}\right)}
{\sin\left(\frac{\pi-6i\beta}{5}\right)}\nonumber \hs\hs .
\end{eqnarray}
Herein $\delta$ is an imaginary number satisfying
\EQ
e^{-2 \pi i \delta}=\frac{s\left(\frac{1}{5}\right)}
{s\left(\frac{2}{5}\right)} \hs\hs .
\EN
All amplitudes but $d(\beta)$ have a simple pole at $\beta=\frac{2\pi i}{3}$.
Their residues are given by
\begin{eqnarray}
\tau_1 & =& \mbox{Res}_{\beta=\frac{2\pi i}{3}} \hs \hs
a(\beta) = i \hs \frac{5}{6} \hs
\frac{\left(s\left(\frac{1}{5}\right)\right)^3}
{\left(s\left(\frac{2}{5}\right)\right)^2} \hs\hs;
\nonumber \\
\tau_2 & =& \mbox{Res}_{\beta=\frac{2\pi i}{3}} \hs
\hs b(\beta) = -i \hs \frac{5}{6} \hs
\frac{\left(s\left(\frac{1}{5}\right)\right)^{\frac{13}{6}}}
{\left(s\left(\frac{2}{5}\right)\right)^{\frac{7}{6}}} \hs\hs;
\nonumber \\
\tau_3 & =& \mbox{Res}_{\beta=\frac{2\pi i}{3}} \hs
\hs c(\beta) = i \hs \frac{5}{6}\hs
\frac{\left(s\left(\frac{1}{5}\right)\right)^{\frac{4}{3}}}
{\left(s\left(\frac{2}{5}\right)\right)^{\frac{1}{3}}} \hs\hs;
\\
\tau_4 & =& \mbox{Res}_{\beta=\frac{2\pi i}{3}} \hs
\hs d(\beta) = 0 \hs\hs;
\nonumber \\
\tau_5 & =& \mbox{Res}_{\beta=\frac{2\pi i}{3}} \hs
\hs e(\beta) = i \hs \frac{5}{6} \hs
\frac{\left(s\left(\frac{1}{5}\right)\right)^{\frac{4}{3}}}
{\left(s\left(\frac{2}{5}\right)\right)^{\frac{1}{3}}} \hs\hs.
\nonumber
\end{eqnarray}
Their numerical values are collected in Table 2.

In the asymptotic limit $\beta \rightarrow \infty$, all amplitudes
but $a(\beta)$ have an oscillating behaviour
\begin{eqnarray}
a(\beta) & \sim & e^{-2\pi i\delta} \hs\hs;\nonumber \\
b(\beta) & \sim & e^{-\delta i \pi}
\hs e^{\delta \beta}\hs e^{\frac{3\pi i}{5}} \hs\hs;\nonumber \\
c(\beta) & \sim & e^{-\delta\beta} \hs e^{-\frac{3\pi i}{5}}\hs\hs;
\label{limite2} \\
d(\beta) & \sim & e^{-2\delta\beta} \hs e^{-\frac{i\pi}{5}} \hs\hs;\nonumber \\
e(\beta) & \sim &
e^{-2\delta i \pi}
\hs e^{2\delta \beta}\hs e^{\frac{-4\pi i}{5}} \hs\hs. \nonumber
\end{eqnarray}
Such oscillating behaviour was also found by the same author for a
scattering model with $Z_4$ symmetry \cite{z4}. There, it was suggested
that in the ultraviolet limit, the $Z_4$ theory has a limit cycle. Whether
or not this is also the case for the $S$-matrix proposed by Zamolodchikov
for the subleading magnetic perturbation of TIM, it is not clear to us
that it is possible to match such oscillating behaviour to either a definite
CFT in the ultraviolet limit, or a choice of generalized statistics for
the kinks.

\resection{Finite-size effects}

The study of the scaling region around a fixed point is simplified
along those directions (in the space of coupling constants) which define an
integrable massive field theory. In such cases, the knowledge of the
theory on mass-shell (the $S$-matrix) makes it possible to characterize
completely the dynamics even off-shell. In particular, using the
Thermodynamical Bethe Ansatz method \cite{TBA} one could compute the
ground-state energy $E_0(R)$ for the theory on a cylinder of width
$R$. If this computation could be performed for the subleading magnetic
perturbation of TIM, it would become easy to decide which of the two proposed
scattering theories is the correct one. Unfortunately, the Bethe-ansatz
technique has not been extented to the case of an $S$-matrix with
kink excitations. Only few examples have been worked out \cite{RTBA1,RTBA2}.
However, we can get around this difficulty using the
``Truncated Conformal Space Approach'' (TCSA) \cite{LMC,YZ,LM} and the
predictions of finite-size theory \cite{sublead,Luscher}.

The TCSA allows us to study the crossover from massless to massive behaviour
in a theory with the space coordinate compactified on a circle of radius $R$.
The method consists in truncating the infinite-dimensional Hilbert space of
the CFT up to a level $\Lambda$ in the Verma modules. Then, the off-critical
Hamiltonian
\EQ
H(\lambda,R)=H_0(R) + V(R) \hs .
\label{Ham}
\EN
is numerically diagonalized. Here, $H_0(R)$ is the Hamiltonian of the fixed
point \cite{Cardy}
\EQ
H_0(R) = \frac{2\pi}{R}\hs \left( L_0 +\overline L_0
-\frac{c}{12}\right) \hs ,
\EN
and $V(R)$ is the interaction term given by the perturbation
\EQ
V(R) = \lambda\hs\int_0^R \Phi_{r,s}(x) \hs dx \hs \hs .
\label{interaction}
\EN
The matrix elements of $V(R)$ are computed in terms of the three-point
functions of the scaling fields of the fixed point. An efficient algorithm has
been developed for performing such computation \cite{LM}. In our
case, the truncation is fixed at level 5 in the Verma modules.
The parameter $\lambda$ in (\ref{interaction}) is a dimensionful coupling
constant, related to the mass scale of the perturbed theory
\EQ
[\lambda] = m^{2-2\Delta_{r,s}} \hs.
\EN
In the following we fix the mass scale by $\lambda=1$.

The energy levels $E_i(R)$ of the Hamiltonian (\ref{Ham}) have the scaling
form
\EQ
E_i(R)=\frac{1}{R} F_i(mR) \hs ,
\EN
with the asymptotic behaviour
\EQ
E_i(R)\simeq\left\{\begin{array}{ll}
\frac{2\pi}{R}\left(2\Delta_i -\frac{c}{12}\right) &
\hs ,\hs mR \ll 1 \hs;\\
\epsilon_0 m^2 R + M_i & \hs , \hs mR\gg 1 \hs \hs .
\end{array}
\right.
\label{asymp}
\EN
Here, $\Delta_i$ is the anomalous dimension of some scaling field in the
ultraviolet regime and $M_i$ is the (multi)particle mass term in the
infrared limit. However, the above infrared asymptotic behaviour holds
only in the ideal situation when the truncation parameter
$\Lambda$ goes to infinity. In practice, for finite $\Lambda$, the linear
behaviour of eq.\hs(\ref{asymp}) is realized only within a finite
region of the $R$ axis. The large $R$ behaviour is dictated by truncation
effects. In order to find the physical regions, we make use of the
following parameter (introduced in ref.\hs\cite{LMC})
\EQ
\rho_i(R)=\frac{d\hs \log E_i(R)}{d\hs \log R} \hs .
\EN
The parameter $\rho_i$ is between the values $\rho_i =-1$ (in the ultraviolet
region) and $\rho_i=1$ (in the infrared one). In the limit of large $R$
(the truncation-dominated regime), $\rho_i=1-2\Delta$.

The ``window'' in $R$ where the linear infrared behaviour holds
depends upon the perturbing field and, for the case of operators with
anomalous dimension $\Delta \geq \frac{1}{2}$, it can be completely shrunk
away. This phenomenon is related to the divergences which appear in a
perturbative expansion of the Hamiltonian (\ref{Ham}), which must be
renormalized. Under these circumstances, it is more convenient to consider
the differences of energies, which are not renormalized.

In the case of the subleading magnetic perturbation of TIM, the anomalous
dimension of $\Phi_{2,1}$ is $\Delta=\frac{7}{16}$, which is near
$\frac{1}{2}$. Looking at fig.\hs (1.a), we see that the onset of the
infrared region of the two lowest levels is around $R\sim 2$ and persists
only for few units in $R$. In this region one can check that they approach
each other exponentially \cite{LMC}
\EQ
E_1-E_0 \sim e^{-m R} \hs ,
\EN
and extract in this way the numerical mass of the kinks
\EQ
m=0.98 \pm 0.02 \hs .
\label{massa}
\EN
{}From fig.\hs6, we see that for the third level, that of one-particle state,
the ultraviolet behaviour extends till $R \sim 0.5$. The crossover region is
in the interval $0.5 \leq R \leq 2$. Beyond this interval, the infrared regime
begins but the ``window'' of infrared behaviour is quite narrow and
is in the vicinity of $R\sim 3$. In such small region, it becomes hard
to extract any sensible result. To overcome this difficulty, it is better to
consider the differences of energies with respect to those of the degenerate
ground states. From fig.\hs(1.b) one can easily read off the mass-gap and see
that is consistent with the value extracted from the exponential approach of
the two lowest levels. In fig.\hs(1.b), the third line defines the threshold,
with a mass-gap $2m$.

In the ideal situation $\Lambda \rightarrow \infty$, the crossover
between the intermediate region ($mR \sim 1$) and the infrared one
($mR \gg 1$) is controlled by off-mass shell effects and has an exponential
behaviour. The computation of these finite-size corrections has been put
forward by L\"{u}scher \cite{Luscher}. In the case of one-particle state,
there are two leading off-mass-shell contributions coming from the
processes shown in fig.\hs7. The first correction involves the on-mass-shell
three-particle vertex $\Gamma$, which is extracted from the residue at
$\beta=\frac{2\pi i}{3}$ of the amplitudes $S_{11}^{11}(\beta)$ (in the case
of RSOS $S$-matrix) and $a(\beta)$ (in the case of Zamolodchikov's
$S$-matrix). The second correction comes from an integral over the
momentum of the intermediate virtual particle, interacting via the
$S$-matrix ($S_{11}^{11}(\beta)$ for the RSOS $S$-matrix
and $a(\beta)$ for that of Zamolodchikov). Therefore we have
\begin{eqnarray}
\Delta E(R) & \equiv & E_2(R)-E_0(R) =m +
i\hs \frac{\sqrt{3} m}{2} \Gamma^2 \exp\left(-\frac{\sqrt{3}
mR}{2}\right) \\
& & -m\hs\int_{-\infty}^{\infty} \frac{d\beta}{2\pi}\hs
e^{-m R \cosh \beta} \cosh \beta \left( S\left(\beta+\frac{i\pi}{2}\right)-1
\right) \hs\hs.\nonumber
\end{eqnarray}
We have done the following. First we have computed numerically the integral on
the intermediate particles in both cases of RSOS and Zamolodchikov's
$S$-matrix and we have subtracted it from the numerical data obtained from
the TCSA. After this subtraction, we have made a fit of the data with a
function of the form
\EQ
G(R)=A+B e^{-\frac{\sqrt{3}}{2} m R} +C e^{-m R}\hs\hs.
\label{fit}
\EN
The first term should correspond to the mass term. The coefficient of
second one is the quantity we need in order to extract the
residue of the $S$-matrix at $\beta=\frac{2\pi i}{3}$
\EQ
\frac{2}{\sqrt{3}m}\hs B=i\hs
\mbox{Res}_{\beta=\frac{2\pi i}{3}} \hs\hs
\left\{ \begin{array}{l}
S_{11}^{11}(\beta)\\
a(\beta)
\end{array}
\right.
\label{resRSOS}
\hs \hs .
\EN
The third term is a subleading correction related to the asymptotic
approach of the lowest levels of our TCSA data to the theoretical vacuum
energy $E_0(R)$.

In the case of RSOS $S$-matrix, the best fit gives the following values
\begin{eqnarray}
A & = & 0.97 \pm 0.02 \hs;\nonumber \\
B & = & -0.29 \pm 0.02 \hs;\label{bestfit1} \\
C & = & -0.36 \pm 0.02 \hs.\nonumber
\end{eqnarray}
The corresponding curve is drawn in fig.\hs8, together with the data
obtained from TCSA. The mass term agrees with our previous calculation
(eq.\hs (\ref{massa})). The second term gives for the residue
at $\beta=\frac{2\pi i}{3}$ the value $0.34 \pm 0.02 $. This is consistent
with that of the RSOS $S$-matrix. In our fit procedure, the value of the
residue we extracted through (\ref{resRSOS}) is stable with respect to small
variation of the mass value. Increasing (decreasing) $m$, $B$ increases
(decreases) as well, in such a way that the residue takes the same value
(into the numerical errors). This a pleasant situation because it permits an
iterative procedure for finding the best fit of the data: one can start
with a trial value for $m$ (let's say $m=1$) and plug it into (\ref{fit}).
{}From the $A$-term which comes out from the fit, one gets a new determination
of the mass $m$ that can be again inserted into (\ref{fit}) and so on.
Continued iteration does not affect significantly the value we extract
for the residue, but converges to an accurate measurement of the mass. The
values in (\ref{bestfit1}) were obtained in this way.

With Zamolodchikov's $S$-matrix the best fit of the data (with
the same iterative procedure as before) gives the result
\begin{eqnarray}
A & = & 0.96 \pm 0.02 \hs;\nonumber \\
B & = & -1.10\pm 0.02 \hs;\label{bestfit2} \\
C & = & 1.14 \pm 0.02 \hs.\nonumber
\end{eqnarray}
The residue extracted from these data ($1.29 \pm 0.01$) is not consistent with
that one of the amplitude $a(\beta)$. The situation does not improve
even if we {\em fix} the coefficient of $e^{-\frac{\sqrt{3}m}{2}}$ to be
that one predicted by Table 2, namely $B=-0.158$ and leave as free parameters
for a best fit $A$ and $C$. In this case, our best determination of $A$
and $C$ were $A=0.965$ and $C=-0.046$. The curve is plotted in fig.\hs8
together with the data obtained from TCSA.

\resection{Conclusions}

The $S$-matrix proposed by Zamolodchikov for the subleading magnetic
perturbation of TIM is a particular solution of the ``Hard Square
Lattice Gas'' \cite{Baxter}, fitted in such a way that it matches the
physical picture coming from the TCSA data, i.e. the presence of two
fundamental kinks and only one bound state thereof. But also the RSOS
$S$-matrix computed in sect.\hs 3 is a particular set of Boltzmann weights for
the ``Hard Square Lattice Gas'' and it reproduces the same features. The key
difference is how the crossing symmetry comes about: for the RSOS
$S$-matrix, one finds a non-trivial charge conjugation matrix,
eq.\hs(\ref{crossing}), whereas for Zamolodchikov's proposal the crossing
symmetry occurs in a standard way, eq.\hs(\ref{cross2}). Their
asymptotic behaviour for large real values of $\beta$ is also quite different:
the RSOS $S$-matrix goes to a definite limit while the $S$-matrix proposed by
Zamolodchikov is oscillating. For the RSOS case, the finite limit of the
$S$-matrix allows us to introduce generalized statistics of the kink
excitations \cite{Swieca,KT,Smirnov2}.

The real discrimination between the two scattering theories proposed for the
subleading magnetic deformation of TIM is seen by comparing them with a
numerical ``experiment'', i.e. from the study of the finite-size corrections
of the energy levels obtained by the Truncated Conformal Space Approach. We
have investigated this problem in sect.\hs 6. The result suggests that
the RSOS $S$-matrix gives a more appropriate description for the scattering
processes of the massive excitations of the model. Though, the interesting
question as to what kind of system the $S$-matrix of Zamolodchikov
corresponds to, remains open.

\section*{Acknowledgments}
We thank T. Miwa, A. Schwimmer and Al.B. Zamolodchikov for useful
discussions. We are grateful to P. Orland for a critical reading of the paper.
One of us (F.C.) thanks P. Di Vecchia for warm hospitality at NORDITA.


\end{document}